# CONSERVATION LAWS IN BIOLOGY AND EVOLUTION, THEIR SINGULARITIES AND BANS

Mark Ya. Azbel',
School of Physics and Astronomy, Tel-Aviv University,
Ramat Aviv, 69978 Tel Aviv, Israel[+]
and
Max-Planck-Institute für Festkorperforschung – CNRS,
F38042 Grenoble Cedex 9, France

Well known biological approximations are universal, i.e. invariant to transformations from one species to another. With no other experimental data, such invariance yields exact conservation (with respect to biological diversity and evolutionary history) laws. The laws predict two alternative universal ways of evolution and physiology; their singularities and bans; a new kind of rapid (compared to lifespan), reversible, and accurate adaptation, which may be directed. The laws agree with all experimental data, but challenge existing theories.

Evolution is dynamics of biological complexity. Mechanisms and regularities, which are common to enormously diverse living beings, may be defined as evolutionary invariants. They are well known on micro (DNA) and mesoscopic (cell) scales, and crucial to genetics and biology at large. On a macroscopic scale metabolism (which allows for entropy decrease), and mortality (which allows for natural selection) are evolutionary and biological must. Their characteristics are known to yield approximate (and rather noisy) universal relations. Empirical allometric relations [1-3] reduce basal (i.e., resting) oxygen consumption rate, heart beat and respiration times, life span to animal mass. Gompertz [4] presented the first mortality law. Thereafter the search for the universal mortality law went on [5,6]. Demographers [7] established regularities and similarities in demographic approximations. However, biological data depend on a multitude of unspecified factors in often little controllable and reproducible conditions [8-11], and empirical universality is sometimes viewed as a mechanistic simplification with little biological insight [12].

In physics exact laws of mechanics were established when friction (which also depends on a multitude of parameters) was disregarded. Similarly, consider only "canonic" fractions of biological characteristics which yield an exact universal law. A relation which is conserved under (invariant to) any transformations from one species to another, is a universal conservation law of biology and evolution. Its very invariance to such an extraordinary wide class of transformations, with no other experimental data, accurately predicts its functional form. The law specifies possible universal ways of evolution, and thus biology of enormously diverse living beings. It unravels a new kind of rapid (compared to lifespan) and reversible directed adaptation, and suggests its mechanism. The law and its predictions are extensively verified. The law is valid for canonic fractions only, but then it is accurate, inconsistent with existing theories, yields new concepts and insights, which challenge common wisdom, and without theoretical approach were previously overlooked.

Start with the derivation of the law. Suppose macroscopic canonic quantities u and v yield an exact universal relation u = f(v). The values of u and v in a population are the averages of their values $u_G$ and $v_G$ in different population groups (e.g., in different living conditions). If the distribution function of $v_G$ in v is $c(v,v_G)$, then

$$\int c(v,v_G) \, dv_G = 1; \qquad v = \int v_G c(v,v_G) dv_G \equiv <v_G> \quad . \tag{1}$$

So, u = f(v) = f(<$v_G$>). Universality implies that $u_G$ = f($v_G$) Suppose u is an additive quantity, i.e. u =<$u_G$>=<f($v_G$)>. Then <f($v_G$)> = f(<$v_G$>), i.e.

$$\int c(v,v_G) \, f(v_G) \, dv_G = f \left[ \int c(v,v_G) \, v_G \, dv_G \right] \tag{2}$$

Equation (2) is a functional equation which is linear in f and non-linear in c. Consider a special case of

$$c = c_1 \delta(v_G - w_1) + c_2 \delta(v_G - w_2), \qquad \text{where } c_1 + c_2 = 1 \tag{3}$$

Then, by Eq. (2), $c_2 f'(w_1) = c_2 f'(w_2)$. Thus, either

$$f(v) = av + b \quad , \tag{4a}$$

where a and b are constants, or $c_2 = 0$, i.e.

$$c(v, v_G) = \delta(v_G - v) \tag{4b}$$

A special case (3) implies that Eqs. (4a, 4b) are necessary for $f(v)$ being a solution to Eq. (2). Since both Eqs. (4a) and (4b) satisfy Eq. (2) in a general case, they are sufficient for $f(v)$ to be a solution to Eq. (2). Thus, Eqs. (4a, 4b) present its two general solutions. Specifically, Eq. (4a) implies the (linear) conservation law but imposes no restrictions on the population heterogeneity; Eq. (4b) implies conserved homogeneity of the canonic population in v, but allows for an arbitrary universal law. Equations (4a) and (4b) present the only general solution with no singularities in $f(v)$. Consider a universal solution to Eq. (2), which has singularities at universal points $v^{(j)}$. Validity of the solution (4a) is related to its linearity and to the normalization condition (1). Singular points $v^{(j)}$ determine successive universal intervals $(v^{(j)}, v^{(j+1)})$. Suppose a canonic population is always distributed inside the same such interval, i.e.,

$$\int_{v^{(j)}}^{v^{(j+1)}} c(v, v_G) dv_G = 1 \tag{5}$$

Then, similar to the previous case, a piecewise linear law

$$u = a^{(j)} v + b^{(j)} \qquad \text{if} \quad v^{(j)} \leq v \leq v^{(j+1)} \tag{6}$$

($a^{(j)}$ and $b^{(j)}$ are universal constants) is a universal solution to Eq. (2). At any intersection $v^{(j)}$, Eq. (5) implies

$$c(v^{(j)}, v_G) = \delta(v_G - v^{(j)}) \quad . \tag{7}$$

Equations (5) and (7) yield a ban (a "heterogeneity exclusion principle") in a canonic population [13] – Eq. (5) excludes segments outside a given interval, and any heterogeneity at the segment boundaries. Since different groups in different conditions are heterogeneous in their sets of factors η, Eq. (7) implies that the susceptibility of v to η vanishes at $v^{(j)}$. Eq. (6) may be presented in the form:

$$u = \xi^{(j)} u^{(j)} + \xi^{(j+1)} u^{(j+1)}; \quad \xi^{(j)} = (v - v^{(j)}) / (v^{(j+1)} - v^{(j)}); \quad \xi^{(j+1)} = 1 - \xi^{(j)} \quad , \tag{8}$$

when
$$v^{(j)} \leq v \leq v^{(j+1)}$$

which maps it onto the coexistence of phases $u^{(j)} = f(v^{(j)})$ and $u^{(j+1)} = f(v^{(j+1)})$ with the "concentrations" $\xi^{(j)}$ and $\xi^{(j+1)}$, and implies homogeneity at the phase boundaries. A single phase implies Eq. (4a); an infinite number of continuous phases implies Eq. (4b) either at any v or at certain v intervals. The latter case implies a singularity at each such interval boundary with the linear law interval (where $f'(v)$ is constant).

Thus, there exist two basic universal ways of "canon" evolution: via linear conservation law (which imposes no restrictions on the population heterogeneity), or via conserved population homogeneity (which allows for any universal law). The crossovers between different laws (linear–linear of linear–nonlinear) imply singularities.

Predicted piecewise linearity is experimentally explicit without any adjustable parameters, and allows for comprehensive verification. Indeed, if in a certain interval a relation between two additive quantities is approximately linear for given populations, then it is also linear for any their mixtures (with the accuracy of the maximal deviation from linearity. One may significantly improve the accuracy of a linear approximation by discarding biologically or experimentally special cases for separate study). Arbitrarily heterogeneous population implies Eq.(1), and thus exact universal linearity between canonic quantities. Verify the predicted conservation law with well known experimental data and empirical approximations. Those between the basal (i.e. at rest) oxygen consumption $v_h$, $v_r$, $v_e$, correspondingly per heartbeat $t_h$, respiration time $t_r$, maximal lifespan $e_m$ and the mass of a warm-blooded animal m are approximately linear [2]. This implies the predicted universal linear law for their canonic fractions $V_h$, $V_r$, $V_m$ vs M (here and on capital letters denote canonic fractions). Oxygen consumption of cold-blooded animals exponentially depends on temperature [1-3]. (Their activation energy is consistent with the hydrogen binding energy [2]). When it is renormalized to 311°K, then it yields the same linear universal law as for warm-blooded animals. Thus, in agreement with Eq. (4a),

$$V_h = a_h M, \qquad V_r = a_r M, \qquad V_m = a_m M . \tag{9}$$

This implies the existence of three universal biological constants (similar to fundamental physical constants, but known with low accuracy): the numbers of consumed oxygen molecules per body atom per lifespan, respiration rate and heartbeat ($\sim 3$; $\sim 10^{-8}$; $\sim 3 \times 10^{-9}$ correspondingly; on their origin see refs. 2, 14). Empirical scaling [1-3] reduces basal oxygen consumption rate $\dot{v}_0$ per unit (i.e. biologically non-specific) time, heartbeat $t_h$ and respiration $t_r$ times, maximal lifespan $e_m$ to an animal mass m:

$$\dot{v}_0 = A_v m^\alpha; \qquad t_h = A_h m^\beta; \qquad t_r = A_r m^\gamma; \qquad e_m = A_m m^\delta \tag{10}$$

where A's and "critical indexes" $\alpha$, $\beta$, $\gamma$, $\delta$ are constants [15]. Non-linear approximately universal scaling implies Eq. (4b) for canonic quantities, and thus their homogeneous constant values. (This is in contrast to Eq. (9) which is valid for any mixed population, e.g., for 2 elephants, 100 humans, 1,000,000 sparrows per their corresponding heartbeat times). Indeed, for a given species all variables (e.g., basal heartbeat time) in Eq. (10) are approximately constant [1]. This poses a physical and biological challenge. The value of $v_G$ in Eq. (4b) depends on a set $\eta$ of different factors [13], and $v_G(\eta) = v$ is constant inside the corresponding multidimensional $\eta$ manifold. An infinitesimally close $v + dv$ corresponds to a different manifold $\eta'$, which is infinitesimally close along a certain area. (E.g., in a one dimensional case $v_G(\eta) = v$ when $\eta_1 \leq \eta \leq \eta_2$, and $v_G(\eta) = v + dv$ when $\eta_2 + d\eta_2 \leq \eta \leq \eta_3$). For a continuous v this implies a singularity of $v_G$ at its $\eta$ boundary, i.e. at every value of v. Finite number of species "quantizes" continuous v into (evolutionary metastable) constants. Singularities at all points and their "quantization" are inconsistent with and a challenge to any theory.

Consider mortality. Start with humans. Their mortality is extensively quantified in demographic "life tables" [7], which use accurately registered human birth and death records. The so-called "period" tables contain the mortality rates $q_x$, i.e., the probabilities to die from age x to x+1 in a given calendar year, for a given sex and country or its specific group (over 50 000 data items for Sweden alone). A "period" survivability $\ell_x$ is the probability to survive to a given age x in a given calendar year. It equals $\ell_x = p_0 p_1 \ldots p_{x-1}$, where $p_y = 1 - q_y$ is the probability to survive from age y to age y+1. "Cohort" life tables list $q_x$ and $\ell_x$ for a "cohort", born the same calendar year. Biodemographic life tables quantify mortality rates and survivabilities, usually for an animal cohort, at its characteristic ages (e.g., days for flies). At any given age empirical relation between $q_x$, $\ell_x$ and $q_0$ (= 1 - $\ell_1$) in protected populations is approximately universal and piecewise linear for species as remote as humans [6,16] and flies [6]. Survivability is additive, since the number of survivors in the population is the sum of their numbers in all population groups. This yields Eq. (2), where u, v are replaced by canonic survivabilities $L_x$ and $L_1$, but f depends also on the "eigentime" x. Consequently, Eq. (6) changes to

$$L_x = a^{(j)}(x) L_1 + b^{(j)}(x) \quad \text{when} \quad L_1^{(j)} \leq L_1 \leq L_1^{(j+1)} \tag{11}$$

Empirical scaling proves that human and fly $a^j(x)$ and $b^j(x)$ reduce to universal functions of age [6]. Thus, the exact piecewise linear law of canonic survivability is biologically non-specific (i.e. independent of genotypes, phenotypes, life history, age specific diseases, and all other relevant factors). At least some deceases, which significantly contribute to mortality (e.g., tuberculosis in pre-1949 Japan and in 1890-1940 Finland), also do not violate the universal law [16]. So, a fraction of their mortality is also canonic. Dependence on x yields new predictions. At any age linear segments intersect at the same values of $L_1^{(j)}$ (thus at the same values of $Q_0^{(j)} = 1 - L_1^{(j)}$). The ultimate boundaries of any probability are 0 and 1. Thus, by Eq. (1), canonic survivability $L_x$ must homogenize there. By Eq. (11), an "initial condition" (at x = 0) $Q_0 = 1 - L_1$ accurately determines canonic mortality rate $Q_x = 1 - L_{x+1}/L_x$ at any age in the same calendar year. Mortality $Q_0$ strongly depends on living conditions, but from conception to x = 1 only. So, at any age canonic mortality rate $Q_x$ rapidly adjusts to, and is determined by, current (< 2 years for humans) living conditions only. It is independent of the previous life history. Therefore, together with $Q_0$, it may be reduced and reversed to its value at a much younger age. So, when mortality of a cohort is predominantly canonic, it may be reversed also. Reversible mortality implies its reversible adjustment to living conditions with the relaxation time small compared to lifespan.

Verify Eq. (11) and its predictions. Equation (11) agrees with empirical piecewise linear law [6,16], which reduces $L_x$ to $L_1$, and whose intersections are indeed simultaneous at all ages. Universal crossovers to different slopes are consistent with significant declines of old age mortality in the second half of the 20[th] century [17]. Demographers interpreted them as 'epidemiological transitions', characterized primarily by the reduction of mortality from cardiovascular diseases. However, the predicted and verified rapid transitions occur simultaneously across generations (which have a different life history behind them and may even be genetically distinguishable). Transitions are universal for humans, med- and fruitflies. This suggests that medical progress just shifts human survivability to a universal transition. Accurate reduction of mortality $Q_x$ at any age to $Q_0$ is amazing (since living conditions, e.g., food and diseases, are intrinsically very different for elderly and newborns). Yet, it is consistent with clinical studies [18], as well as with demographic observation that infant mortality is a sensitive barometer of mortality at any age [7]. However, only exact universal law accurately predicts mortality reversibility, which is inconsistent with any evolutionary theory of aging [9]. Remarkably, this crucial prediction agrees with demographic data. For instance, mortality rate of Swedish females, born in 1916, at 48 years returned to its value at 20 years. Human survivability at any age extrapolates [16] to 1, suggesting that canonic mortality may be entirely eliminated. Thus, total mortality may be significantly decreases, and life expectancy significantly increased, in agreement with ref. 19 and other demographic data. In the last 30 years (1965-1995) Japanese females almost halved their mortality at 90 years, and increased their period probability to survive from 60 to 90 years 4.5-fold, to remarkable 33% of survivors.

Consider homogenization of Swedish female and male survivabilities (which depend on both genetic and environmental factors), and their vanishing susceptibility to living conditions, at the crossovers. In agreement with the prediction, their relative difference decreases 5-fold to a minimum in the vicinity of the main crossover, then reaches a maximum, and finally decreases to $\ell_1 = 1$ (see Fig. 1) when the population infant mortality monotonically decreases fifty fold from 1861 till 1999. Such non-monotonic change in steadily improving conditions was unanticipated in any of multiple theories[9] which relate mortality to mutation accumulation and cumulative damage; telomere; oxygen consumption; free radicals; life-history trade-off; relation between reproductive rate and nutrient supply; lethal side-effect of a late-onset genetic disease [20]. In contrast to these theories, universal conservation law suggests the existence of a new unusual mechanism of mortality, which allows for rapid reversible adjustment to changing living conditions, in particular via singularities and population homogenization, and dominates in evolutionary unprecedented protected populations. The only known reversible processes in a macroscopic system are adiabatic changes in its equilibrium state. Exact universal (i.e., biologically non-specific, independent of genotypes, phenotypes, life history, age specific diseases, and all other relevant factors) law is characteristic for physics rather then biology. Thus, its accurate mapping (8) onto phase equilibrium, its singularities, and its number of variables (unlike multi-variable potential in the conservation law of mechanical energy) may not be a coincidence. All this calls for comprehensive verification of the universal law in quantitatively controllable conditions.

To summarize. A conservation law, i.e. an exact relation between certain fractions of biological quantities, is derived. On a long time scale it proves the existence of universal biological and evolutionary constants, similar to fundamental constants in physics. They change via universal singularities, where susceptibility to different conditions vanishes. On a short time scale the law challenges any existing theory with predictions of "quantized" species specific constants, which slowly and universally change on a larger time scale, and a new kind of reversible adaptation, its singularities and their bans on population heterogeneity.

Acknowledgement. Financial support from A. von Humboldt award and R. & J. Meyerhoff chair is appreciated. I am very grateful to I. Kolodnaya for technical assistance.

[+] Permanent address

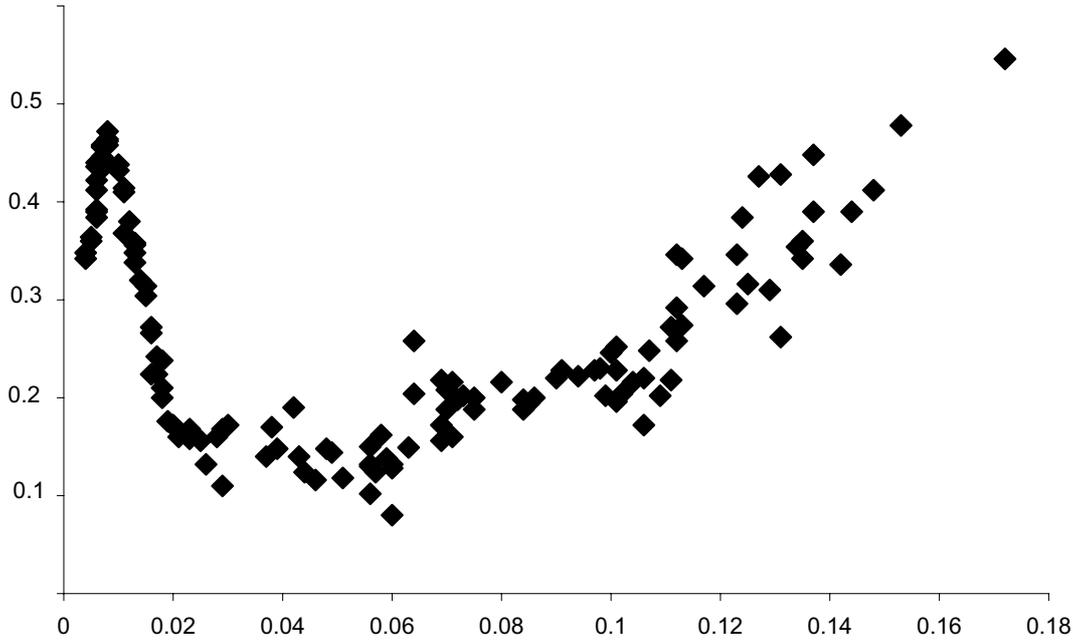

Fig. 1. Relative difference $\delta_{80} = \Delta\ell_{80} / \overline{\ell}_{80}$ vs $\overline{q}_0$ for 1861–1999 Sweden. Here $\Delta\ell_{80}$ is the difference of female and male survivabilities at 80 years in the same calendar year; $\overline{\ell}_{80}$ is the survivability at 80 years and $\overline{q}_0$ is the infant mortality of an entire population (which changes 50-fold).